\documentclass[showpacs,twocolumn,floatfix,amsmath,amssymb,superscriptaddress,prl]{revtex4-1}
\usepackage{epsfig}
\usepackage{graphicx}
\usepackage{longtable}
\usepackage{graphicx}
\usepackage{dcolumn}
\usepackage{multirow}
\usepackage{bm}
\usepackage{ulem}
\usepackage{color}
\usepackage{braket}

\begin{document}

\title{Topological end states due to inhomogeneous strains in \\ wrinkled semiconducting ribbons }

\author{Sudhakar Pandey}
\affiliation{Institute for Theoretical Solid State Physics, IFW Dresden, Helmholtzstr. 20, 01171 Dresden, Germany}

\author{Carmine Ortix}
\affiliation{Institute for Theoretical Solid State Physics, IFW Dresden, Helmholtzstr. 20, 01171 Dresden, Germany}
\affiliation{Institute for Theoretical Physics, Center for Extreme Matter and Emergent Phenomena, Utrecht University, Leuvenlaan 4, 3584 CE Utrecht, Netherlands}

\begin{abstract}
We show that curvature-induced inhomogeneous strain distributions in nanoscale buckled semiconducting ribbons lead to the existence of end states which are topologically protected by  inversion symmetry. These end-state doublets, corresponding to the so-called Maue-Shockley states, are robust against weak disorder. By identifying and calculating the corresponding topological invariants, we further show that a buckled semiconducting ribbon undergoes topological phase transitions between trivial and non-trivial insulating phases by varying its real space geometry. 
\end{abstract}

\pacs{73.22.-f, 03.65.Vf, 73.21.Cd, 73.20.At}

\maketitle

\paragraph {Introduction.--- } In recent years, the investigation of topological states of matter has become a subject of growing interest \cite{has10,moo10,qi11}. It has brought forth the theoretical prediction \cite{kan05,kan05b,ber06,fu07,fu11} and experimental verification \cite{kon07,hsi09,ras13,pau15} of a plethora of different topologically nontrivial electronic quantum phases. Contrary to their trivial counterparts, topologically nontrivial quantum phases exhibit protected surface or edge states that lie inside the bulk gap. These topological states are a direct physical consequence of the topology of the bulk band structure, which is characterized by a quantized topological invariant \cite{ryu10}. 
It is fair to say that low-dimensional semiconducting nanomaterials are the primary solid-state setups where signatures of topological states of matter have been uncovered. The quantum spin Hall effect, for instance, has been experimentally verified in HgTe quantum wells \cite{kon07}, while signatures of Majorana bound states have been reported in a heterostructure comprising a semiconductor nanowire with strong spin-orbit coupling and a conventional $s$-wave superconductor\cite{mou12}. 

The rapid advances in synthesizing  low-dimensional nanostructures in which semiconductor nanomaterials are bent into curved, deformable objects such as spiral-like nanotubes \cite{pri00,sch01}, nanohelices \cite{zha02} and even more complex three-dimensional nanoarchitectures \cite{xu15}, provides us a whole new family of solid-state platforms where topological non-trivial states of matter can arise. These geometrically deformed nanostructures display a variety of unique curvature-induced phenomena, which include, among others, bound states formation \cite{ort10} and a strongly directional dependent ballistic magnetoresistance \cite{cha14}. In addition, it has been recently shown that curvature effects in a bent nanowire with Rashba spin-orbit interaction can promote the generation of topological insulating phases in an otherwise metallic system \cite{gen15}. 

\begin{figure}
\includegraphics[width=\columnwidth]{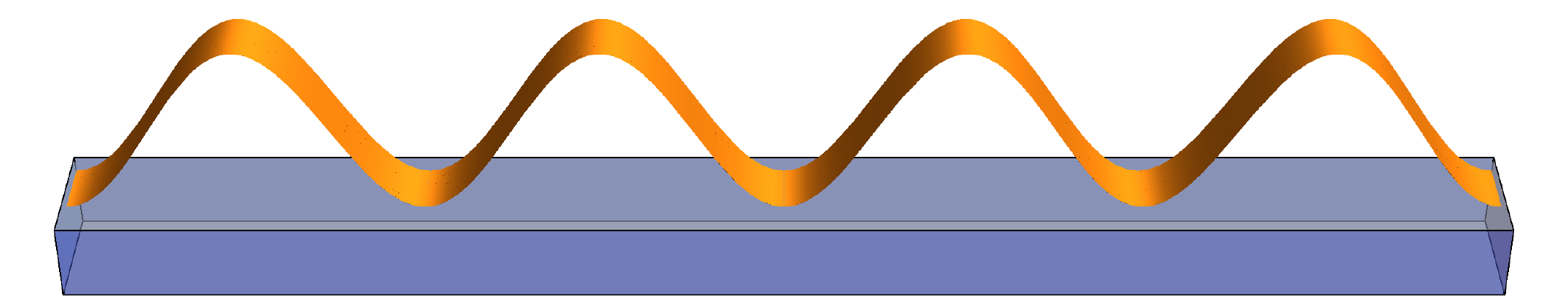}
\caption{(color online) Schematic of a wrinkled semiconducting nanoribbon on an elastomeric substrate with litographically patterned surface adhesion sites.}  
\label{fig:fig0}
\end{figure}

Motivated by the possibility of creating one-dimensional ``wavy" nanostructures [c.f. Fig.~\ref{fig:fig0}], as obtained by depositing semiconducting nanoribbons on elastomeric prestrained substrates with patterned surface adhesion sites \cite{sun06,rog08},
in this Letter  we prove that in materials with negligible spin-orbit coupling, curvature-induced inhomogeneous strain distributions lead to the appearance of localized in-gap states -- the so-called Maue-Shockley bound states \cite{mau35,sho39} --  topologically protected by a one-dimensional inversion symmetry. These bound states are stable against weak disorder and eventually result in an effective double quantum dot system. By  identifying and calculating the corresponding bulk topological invariant, we further show the occurrence of topological phase transitions while changing the real space geometry, thereby signalling a strong interconnection between the local curvature and the topology of the electronic states. 

\paragraph{Effective ${\bf k \cdot p}$ model. ---} Our starting point is the effective ${\bf k \cdot p}$ Hamiltonian for conduction electrons in a semiconducting ribbon shaped in a wrinkled structure with parametric equations ${\bf r} = {\bf r}(s)$, where $s$ indicates the arclength of the planar curve measured from an arbitrary reference point.  A key property of any generic bent nanostructure is the nanoscale variation of strain,  the dominant component of which varies linearly across the thickness as $\epsilon_{ss} = - q_{3} \kappa(s)$, where $\kappa(s)$ indicates the local curvature while $q_3=0$ defines the mechanical neutral plane \cite{lan86}. Taking into account the strain-induced shifts of the conduction band edges \cite{wal89}, it has been shown \cite{ort11} that this inhomogeneous strain distribution induced by curvature yields a strain-induced geometric potential (SGP) that is of the same functional form as the curvature-induced quantum geometric potential (QGP) introduced by Jensen, Koppe and Da Costa \cite{jen71,dac81}, but (strongly) boosting it. As a result, the effective one-dimensional Schr\"odinger equation \cite{ort15} for a bent nanoribbon reads \cite{note}: 
\begin{equation} 
-\dfrac{\hbar^2}{2 m^{\star}} \left[\partial_s^2 + \dfrac{\kappa(s)^2}{4} \times v_R  \right] \psi = E \psi , 
\label{eq:schreq}
\end{equation}
where $v_R \geq 1$ is a strain-induced renormalization of the QGP whose value depends on the nanoribbon thickness $\delta$ and the material specific deformation potential. 
To proceed further, we make use of the fact that the geometry of wrinkled ribbons can be fairly approximated \cite{rog08} with a simple sinusoidal form $y(x)=A \sin{q x}$  where $A$ corresponds to the maximum wrinkle height while $ 2 \pi / q$ is the corrugation period. Since the curvature is periodic in this geometry, it then follows that the SGP renders an effective superlattice potential, which in the shallow deformation limit $A q \ll 1$ corresponds to a simple harmonic potential ${\cal U}(s) = {\cal U}_0 \cos{2 q s}$ of ${\cal U}_0=  \hbar^2 A^2 q^4  v_R / (16 m^{\star})$ strength and $\pi/q$ period. This also implies the opening of a gap at the mini Brillouin zone (mBZ) edges $k = \pm q$ ($k$ indicating the momentum along the tangential direction of the wrinkled ribbon), thereby leading to a metal-insulator transition  for carrier densities such that the Fermi energy lies in the miniband gap.

Besides this metal-insulator transition, the presence of a curvature-induced superlattice potential also points to the possible occurrence of in-gap  localized end modes, corresponding either to the so-called  Maue-Shockley (MS) \cite{mau35,sho39} or to the Tamm-Goodwin \cite{tam32,goo39} end states  depending on whether the wrinkled ribbon is terminated at the inversion centers of the superlattice potential or not. We will concentrate on the former situation since the occurrence of in-gap end modes follows a topological criterion, which we will demonstrate to be equivalent to the symmetry criterion introduced by Zak \cite{zak85}. The effective Hamiltonian Eq.\ref{eq:schreq} for a wrinkled ribbon possesses time-reversal symmetry ${\mathcal T}$ (${\mathcal T}^2 = 1$) and an inversion symmetry ${\mathcal P}$ (${\mathcal P}^2=1$) with the inversion centers being either at the points of zero curvature $\kappa(s) \equiv 0$ or at the points of maximum curvature, {\it i.e.} at the valleys and crests of the wrinkles. Since $\left[ {\mathcal P}, {\mathcal T} \right] \equiv 0$, the effective Hamiltonian for a wrinkled ribbon falls into the inversion symmetric orthogonal class (AI)  of topological insulators with additional point group symmetries introduced by Lu and Lee \cite{lu14}, who thereby extended the famous Altland-Zirnbauer (AZ) table \cite{alt97,sch08}. Specifically, while the class AI of the original AZ table is trivial in one-dimensional systems, the additional inversion symmetry allows for a ${\mathbb Z}$ topological invariant, which can be computed as follows.  

Considering a centrosymmetric unit cell of the superlattice, the Bloch Hamiltonian ${\cal H}(k)$ commutes with parity operator ${\mathcal P}$ at the center and at the edges of the mBZ. The fact that the eigenstates of the Bloch Hamiltonian have a well-defined parity $\zeta_i$, with $i$ the band index,  at these inversion-symmetric momenta allows to define an integer invariant \cite{hug11,lau15}: 
\begin{equation}
{\cal N} :=  \lvert n_1-n_2 \rvert
,\label{eq:tinv}
\end{equation}
where $n_1$ and $n_2$ are the number of the negative parity occupied eigenstates at $k=0$ and $k=q$ respectively. By first considering the inversion center at the zero curvature point, we find a non-trivial integer invariant ${\cal N}=1$. On the contrary, the integer invariant ${\cal N}=0$  [see Table \ref{tab:tab1}] with the inversion center at the points of maximum curvature, as follows from the fact that a translation of one-half of the superlattice vector switches the parity of the eigenstate at the edge of the mBZ. 
This means that a long finite wrinkled ribbon centered at a zero curvature point and terminated at a crest (valley) of the wrinkle, where the SGP assumes its minimum value,  will generally display a pair of degenerate end modes if the Fermi energy lies in the miniband gap, whereas a finite ribbon centered at a crest or a valley of the wrinkle and terminated at  zero curvature points will manifest its topological triviality with the generic absence of an end-mode doublet. 
\begin{table}[t]
\centering
\begin{ruledtabular}
\begin{tabular}{clccc}
 \multicolumn{2}{c}{IC} & $\zeta_1 (k=0)$  & $\zeta_1(k=q)$ & $ {\cal N} $ \\  [1ex]
\hline 
$\kappa \equiv 0$  && $1$ & $-1$ & $1$  \\ [1ex] 
$|\kappa|_{max}$ && $1$ & $1$ & $0$ \\
\end{tabular}
\end{ruledtabular}
 \caption{Parities of the eigenfunctions at the inversion-symmetric momenta in the first miniband of a wrinkled ribbon considering the shallow deformation limit $A q \ll 1$. The parities have been determined by exact diagonalization of Eq.~\ref{eq:schreq} with the simplified form of the SGP ${\cal U}(s) = {\cal U}_0 \cos{2 q s}$ considering the two inequivalent inversion centers (IC).}
\label{tab:tab1}
\end{table}
This result is in agreement with the existence criterion \cite{zak85} of MS end modes in a one-dimensional crystalline periodic potential which simply states that in-gap states are encountered if the  potential assumes negative values at symmetry centres boundaries.

\paragraph{Topological end-modes. -- }  In order to explicitly prove the appearance of end-modes in the topological non-trivial phase, we next introduce a tight-binding model obtained by discretizing the effective ${\bf k \cdot p}$ model on a lattice. It can be written as: 
\begin{equation}
{\cal H} =  -  t \sum_{j} (c_j^\dagger c_{j+1} + h.c.) + 
\sum_{i}  [ 2t + {\cal U}(s_j) ] c_j^\dagger c_j ,
\label{eq:tightbinding}
\end{equation}
where $c_j^\dagger, c_j$ are operators creating and annihilating, respectively, an electron at the $j$-th site, $t={\hbar^2}/(2 m^{\star}a^2)$ is the hopping amplitude (with $a$ being the lattice constant), and 
${\cal U}(s_j)= {\cal U}_0 \cos[2 q s_j + \vartheta]$, where the atomic positions $s_j = j a$ and $\vartheta$ accounts for non-equivalent displacements of the atoms in one superlattice period. 
In addition we will restrict to the simplest situation for which inversion symmetry can be realized, {\it i.e.} 
$q= \pi / (4 a)$. 
We emphasize that this model has been recently employed to describe end states in a quantum wire with an additional external gate-induced potential having a mirror symmetric profile ($\vartheta=\pi/2$) \cite{gan12}. For our analysis, however, it is essential to consider centrosymmetric unit cells corresponding to $\vartheta=3 \pi /4, -\pi/4$ with the parity operator ${\cal P} = \sigma_x \otimes \sigma_x$.  By calculating the topological invariant of Eq.~\ref{eq:tinv}, we find that the insulating phase at $1/4$ filling is topologically non-trivial for $\vartheta=3 \pi /4$  while a trivial insulating state results for $\vartheta=-\pi/4$. 
Fig.2(a),(b) shows the ensuing energy spectrum with open boundary conditions and an integer number of unit cells. As expected, in the topologically non-trivial insulating phase we clearly find a degenerate doublet of in-gap states localized at the left and the right boundaries of the atomic chain. 

\begin{figure}
\includegraphics[width=\columnwidth,height=.45\columnwidth]{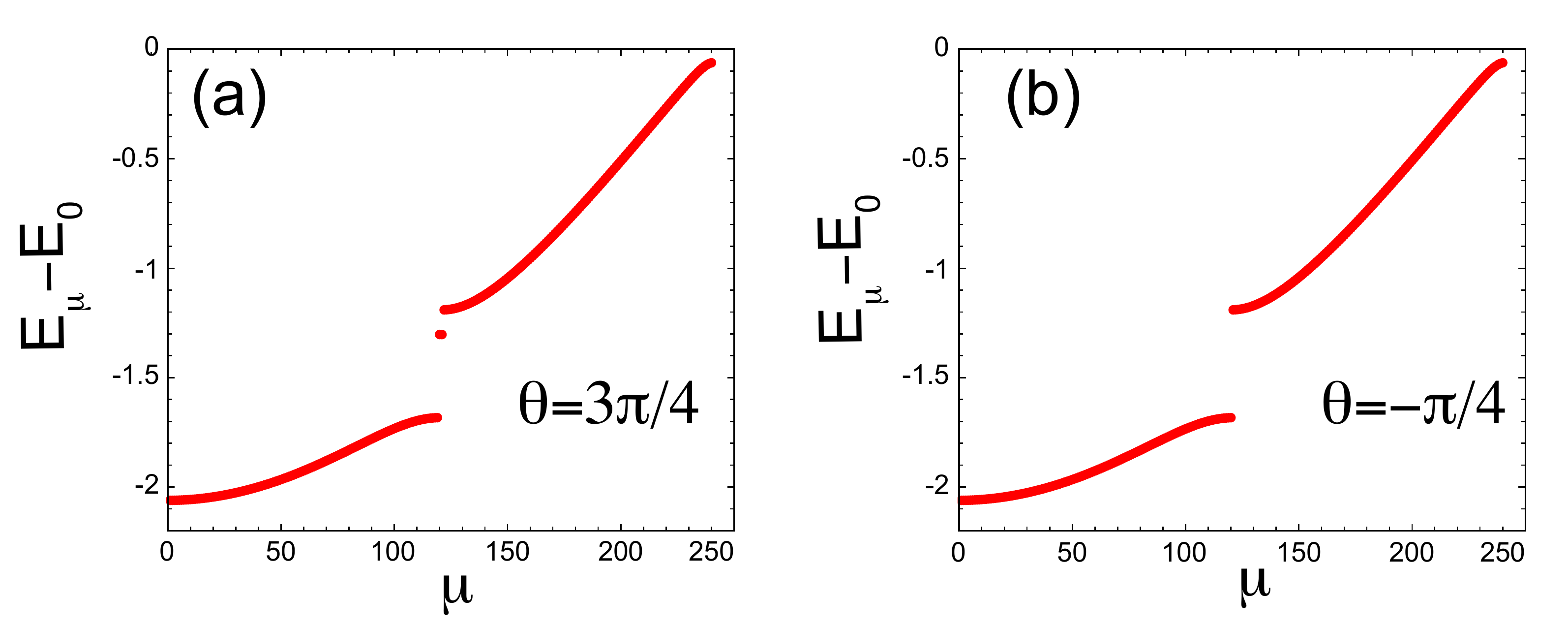}
\caption{(color online) Energy spectrum of the model Hamiltonian Eq.~\ref{eq:tightbinding} for ${\cal U}_0 = t / 2$ and $\vartheta=3 \pi /4$ (a), $\vartheta=-\pi/4$ (b). The spectrum has been obtained by exact diagonalization for a finite atomic chain containing $N=480$ atoms and open boundary conditions. We only show the part of the spectra around the lowest energy gap. Energies have been measured in units of $t$ from a reference energy $E_0$.}  
\label{fig:fig1}
\end{figure}

\begin{figure}
\includegraphics[width=.9\columnwidth]{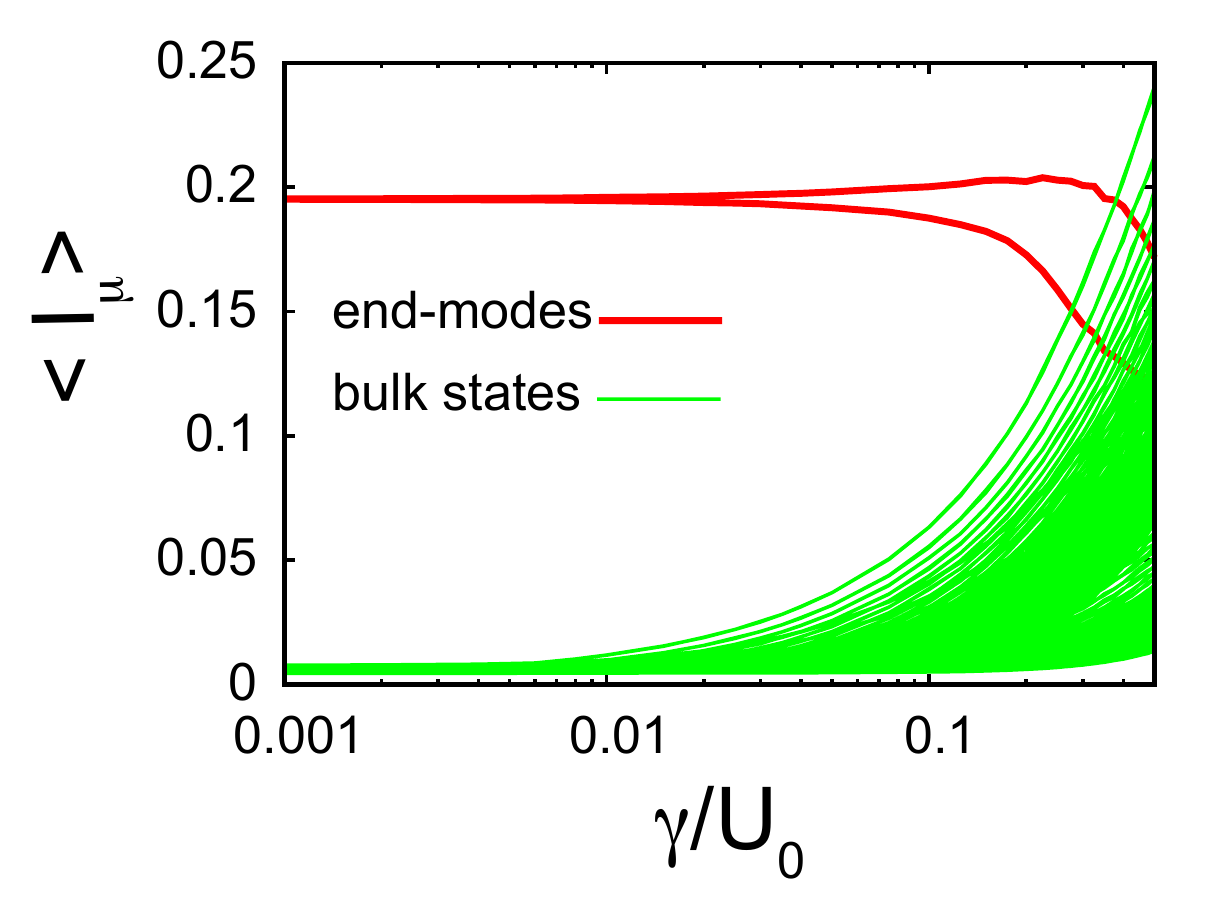}
\caption{(color online) The dependence of the the disorder-averaged IPR $< I_{\mu} >$ on the standard deviation of the random disorder potential for a $N=320$ site chain with potential strength ${\cal U}_0 = t / 2$.}  
\label{fig:fig2}
\end{figure}

Since the topological properties are mandated by a point-group symmetry, 
the end-mode doublets are generally not protected against disorder. 
We have therefore studied this effect by adding a random on-site potential $\sum_j  W_j c_j^\dagger c_j$ where 
$W_j$ is taken according to a Gaussian distribution with zero mean and standard deviation $\gamma$. 
We analyze the fate of the end-modes by looking at the inverse participation ratio (IPR) since it provides a direct and quantitative measure of localization. The IPR of a given eigenvector $\ket{\mu}$ is defined as $I_{\mu} = \sum_{j=1}^N \left( \braket{j | \mu} \right)^4$ where $\ket{j}$ is the canonical site basis \cite{met10}. The IPR is thus restricted to the interval $0 \leq I_\mu \leq 1$, with states perfectly localized on a single site satisfying $I_{\mu} \equiv 1$. Fig.~\ref{fig:fig2} shows the behavior of the disorder-averaged IPR \cite{met13} of a N=320 site chain for the occupied states at $1/4$ filling  and $\theta=3 \pi /4$. For very weak disorder ($\gamma / {\cal U}_0 <0.05$) the IPR of the bulk states displays a substantial increase due to the onset of Anderson localization.  However, the IPR of the end-mode doublets remains unaffected in this regime signalling their stability against mild disorder. As $\gamma$ is further increased, the end states start to mix with the other nearby localized states leading to an effective delocalization and thus a decrease in the IPR.

\paragraph{Topological phase transitions. -- } The functional form of the SGP in the shallow deformation limit corresponding to a simple harmonic potential implies that the results above apply to a number of setups, which include but are not limited to, the aforementioned quantum wires with a gate-induced modulated potential \cite{gan12} or hard-core bosons in one-dimensional optical superlattices \cite{mar15}. 
We next prove that the specific functional form of the SGP leads, in a very natural way, to topological phase transitions between insulating phases with different ${\mathbb Z}$ invariants,  not encountered in other  solid-state setups. Moreover, since these topological phase transitions can be engineered by continuously changing geometrical parameters only, our results will demonstrate that  the topology of the electronic states is intertwined with the real space geometry of low-dimensional curved nanostructures \cite{gen15}. 

\begin{figure}
\includegraphics[width=.9\columnwidth]{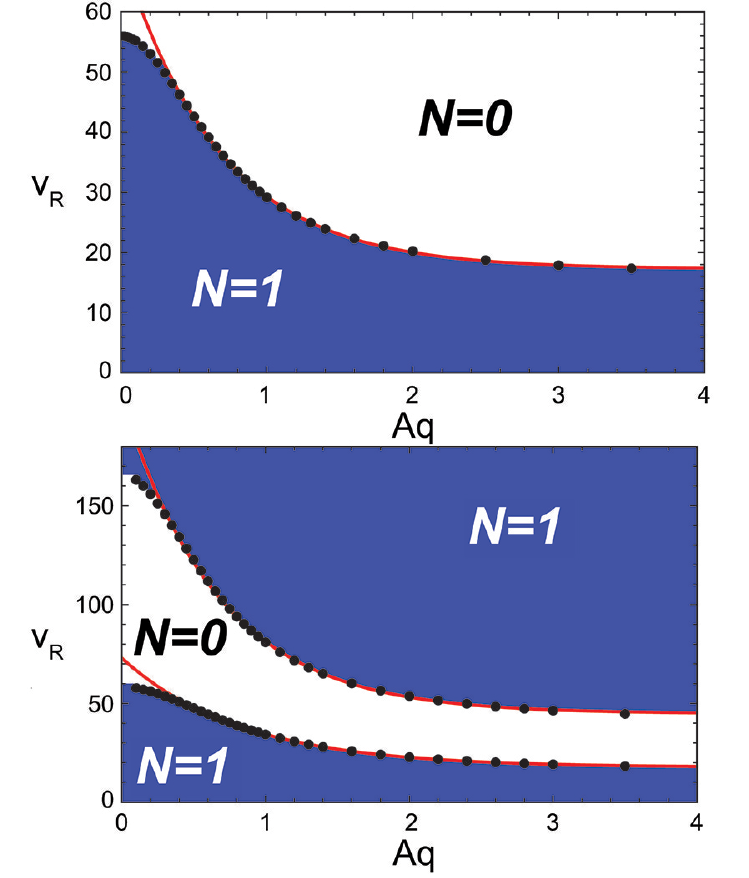}
\caption{(color online) Topological phase diagrams for the existence of localized end states in the second miniband gap (top panel) and the third miniband gap (bottom panel). The dots correspond to the topological phase transition points as obtained by exact diagonalization of Eq.~\ref{eq:schreq}, while the continuous lines are fits as explained in the text.}  
\label{fig:fig3}
\end{figure}

To show the point above, we resort to the continuum ${\bf k \cdot p}$ model and relax the shallow deformation assumption $A q \ll 1$. This can be accomplished by Fourier decomposing the SGP using the functional forms of the arclength $s=\int^x \sqrt{1+ y^{\prime}(x)^2}$  and of the local geometric curvature $\kappa(x) = y^{\prime \prime} (x) /[1 + y^{\prime \,2}]^{3/2}$. By increasing the wrinkle height, the SGP acquires an increasingly anharmonic profile thereby leading to the opening of sizeable gaps at higher energies. However, by sweeping the strain-induced renormalization factor $v_R$, which, as mentioned above, is controlled by the ribbon thickness, we consistently find that the higher miniband gaps undergo closing-reopening processes. Specifically, the gap between the $i$-th and the $(i+1)$-th miniband displays an $(i-1)$ number of closing-reopening points as $v_R$ is increased. 

We illustrate the gap closing-reopening mechanism by particularizing to the gap between the second and third miniband at the mBZ centre. We employ quasi-degenerate perturbation theory, according to which the resulting gap behaves as $\Delta \sim \left| {\cal U}_{2 G} + \alpha \,  {\cal U}_{G}^2 \right|$, where the two dominant Fourier amplitudes of the SGP ${\cal U}_{G,2G}$ depends linearly on the strain-induced renormalization factor $v_R$. The fact that consecutive Fourier amplitudes of the SGP are opposite in sign immediately yields the existence of a critical strain-induced renormalization factor where $\Delta \equiv 0$. The top panel of Fig.~\ref{fig:fig3} shows the behavior of this gap closing-reopening point in the $A q - v_R$ plane. For large deformations ($ A q > 1$) we find that the critical renormalization factor takes the functional form $v_R^c = v_R^0 + \alpha \exp{( - \beta A q)}$ [c.f. the continuous line in the top panel of Fig.~\ref{fig:fig3}] with $v_R^0, \alpha, \beta$ constants. We have verified the exponential behavior of the critical renormalization factor $v_R$ to occur also for the closing-reopening points of the third gap at the mBZ edge [c.f. bottom panel of Fig.~\ref{fig:fig3}].

Since the closing and reopening of a gap is generally accompanied by a topological phase transition, we have computed the integer topological invariant of Eq.~\ref{eq:tinv} and have indeed verified that eigenstates switch their parity at the gap closing-reopening points. In Fig.~\ref{fig:fig3} we show the topological invariants for the second and third gap considering the inversion center at the point of zero curvature. We emphasize that the topological invariants considering the inversion center at the points of maximum curvature will be interchanged for the odd-numbered gaps whereas the topological invariants associated to the even-numbered gaps do not depend on the choice of the inversion center. 
Most importantly we find that in the ultrathin limit $v_R \rightarrow 1$, where curvature effects of quantum origin only survive, a buckled semiconducting ribbon terminated at the crests or valleys of the wrinkles will display topological end modes at all miniband gaps. On the contrary, large strain-induced effects ($v_R \rightarrow \infty$)  trivialize the insulating states at the even-numbered gaps while leaving topological insulating states at the odd-numbered gaps. 

\paragraph{Conclusions --} To conclude, we have shown that in wrinkled semiconducting nanoribbons, inhomogeneous strain distributions induced by curvature lead to the occurrence of localized end states, known as Maue-Shockley bound states, topologically protected by  one-dimensional inversion symmetry.  Their energies lie inside the curvature-induced miniband gaps and are protected by the continuum. The presence of these localized end states can be experimentally detected via different experimental techniques, such as tunneling density of states. 
For a GaAs nanoribbon ($m^{\star} = 0.067 m_e$) with $1 \mu$m wrinkle period and a typical strain-induced renormalization factor $v_R \simeq 10^4$ \cite{ort11}, the characteristic strength of the SGP 
${\cal U}_0 \simeq 0.5 $ meV. Therefore, the upper bound for detecting the topological edge states is for temperatures in the achievable range of a few Kelvin. 
Our results thereby suggest the potential use of wrinkled semiconducting ribbons as an effective double dot system \cite{gan12}, which can be potentially used to implement quantum computing gates for spin qubits \cite{los98}.

We thank Alexander Lau for stimulating and fruitful discussions, and acknowledge the financial support of the Future and Emerging Technologies (FET) program within the Seventh Framework Programme for Research of the European Commission, under FET-Open Grant No. 618083 (the CNTQC project). C.O. thanks the Deutsche Forschungsgemeinschaft (Grant No. OR 404/1-1) for support.


\begin{thebibliography}{9}
\bibitem{note} 
For simplicity we omit the spin indexes, since the two spin channels are independent leading to an additional  twofold degeneracy in the spectrum.
\end{thebibliography}
\end{document}